\documentclass[
aps,
 amsmath,
 amssymb,
reprint,
superscriptaddress,
floatfix,
nofootinbib,
]{revtex4-1}

\usepackage{graphicx}
\usepackage{dcolumn}
\usepackage{bm}

\usepackage{xcolor}
\usepackage{booktabs}
\usepackage{wasysym}
\usepackage[uncertainty-mode = separate]{siunitx}
\usepackage[utf8]{inputenc}
\usepackage[T1]{fontenc}
\usepackage{etoolbox}
\usepackage{hyperref}
\usepackage{stmaryrd}

\hypersetup{
    colorlinks = true,
    linkcolor = blue,
    citecolor = blue,
    urlcolor=blue,
}

\newcommand{\blue}[1]{\textcolor{blue}{#1}} 

\usepackage{tabularx}
\usepackage{appendix}
\usepackage{caption}
\makeatletter
\def\@email#1#2{%
 \endgroup
 \patchcmd{\titleblock@produce}
  {\frontmatter@RRAPformat}
  {\frontmatter@RRAPformat{\produce@RRAP{*#1\href{mailto:#2}{#2}}}\frontmatter@RRAPformat}
  {}{}
}%
\makeatother

\usepackage[font=small,labelfont=bf,format=plain]{caption}

\usepackage{setspace}
  \usepackage{parskip}
\usepackage[compact]{titlesec}         
\titlespacing{\section}{4pt}{4pt}{4pt} 
\AtBeginDocument{
  \setlength\abovedisplayskip{4pt}
  \setlength\belowdisplayskip{4pt}
  }
  
\usepackage[]{caption, subfig}
\usepackage{lineno}
\setcitestyle{square}
\usepackage{newunicodechar}
\newunicodechar{̈}{"{}}

\begin{document}

\newcommand{\ra}[1]{\renewcommand{\arraystretch}{#1}}
\newcommand{\deriv}{\mathrm{d}}

\title{Modeling runaway electron mitigation via passively driven plasma instabilities in a high-current tokamak}
\author{R. Datta}
\thanks{rdatta@mit.edu}
\affiliation{Plasma Science and Fusion Center, Massachusetts Institute of Technology, Cambridge, MA 02139, USA\looseness=-10000 
}%

\author{C. Clauser}
\affiliation{Plasma Science and Fusion Center, Massachusetts Institute of Technology, Cambridge, MA 02139, USA\looseness=-10000 
}%

\author{G. Hu}
\affiliation{Plasma Science and Fusion Center, Massachusetts Institute of Technology, Cambridge, MA 02139, USA\looseness=-10000 
}%

\author{A. Braun}
\affiliation{Columbia University, New York City, NY 10032, USA\looseness=-10000 
}%

\author{N. Ferraro}
\affiliation{Princeton Plasma Physics Laboratory, P.O. Box 451, Princeton, NJ 08543-0451, USA\looseness=-10000 
}%

\author{R. Sweeney}
\affiliation{Commonwealth Fusion Systems, Devens, MA 01434, USA\looseness=-10000 
}%

\author{R. A. Tinguely}
\affiliation{Plasma Science and Fusion Center, Massachusetts Institute of Technology, Cambridge, MA 02139, USA\looseness=-10000 
}%



\begin{abstract}

A critical challenge for tokamak-based fusion power plants is the formation of high-current, relativistic runaway electron (RE) beams during abrupt confinement loss events called disruptions. RE beams may significantly damage machine components if left unmitigated, jeopardizing the operational and economic viability of fusion power plants. We present the first self-consistent extended magnetohydrodynamic (MHD) computational study of RE mitigation with a passively driven RE mitigation coil (REMC) in a high-current tokamak. RE generation and transport are coupled with REMC-driven MHD activity in a 3D simulation that models the thermal and current quench phases of a mid-plane disruption, together with the vertical displacement and final termination of the plasma and RE beam on SPARC – a high-field, high-current tokamak designed to achieve a fusion gain $Q>1$. We demonstrate that REMC-driven magnetic stochasticity can reduce the RE current by over an order-of-magnitude when compared to unmitigated disruptions. The final termination is mediated via a post-thermal-quench "cold" vertical displacement of the plasma, which, together with magnetic stochasticity, increases RE losses via scrape-off once the plasma contacts the first wall. These results demonstrate a viable mechanism for RE mitigation using passively driven plasma instabilities on high-current tokamaks.

\end{abstract}

\maketitle

High-current tokamaks, which magnetically confine a plasma fuel in a toroidal configuration, represent a crucial and promising next step in the quest for commercial fusion energy \citep{barbarino2020brief}. However, future tokamaks, such as SPARC \citep{sweeney2020mhd}, ARC \citep{sweeney2026arc}, ITER and DEMO \citep{lehnen2015disruptions,boozer2017runaway,boozer2018pivotal,boozer2019magnetic}, which will have plasma currents in the range $I_p {\sim}\SI{8}{}\text{-}\SI{20}{\mega\ampere}$, are susceptible to the formation of potentially damaging, mega-Ampere-level runaway electron (RE) beams during disruptions. This is because the multiplication of a RE seed via avalanching is exponentially sensitive to the pre-disruption plasma current, placing RE generation on power-plant-scale tokamaks in a new {\it avalanche-dominated} regime, typically inaccessible on present-day devices \citep{martin2015avalanche,breizman2019physics,boozer2019magnetic}. RE mitigation remains a critical, yet unsolved, challenge for fusion power-plant-scale tokamaks, where RE strikes on plasma-facing components (PFCs) may generate large, localized heat fluxes, causing melting, vaporization, or brittle failure of PFCs \citep{rizzi2026thermal,ratynskaia2026thermal}, as seen on various existing machines \citep{ratynskaia2026roadmap}.

\begin{figure*}[t!]
\includegraphics[page=1,width=1.0\textwidth]{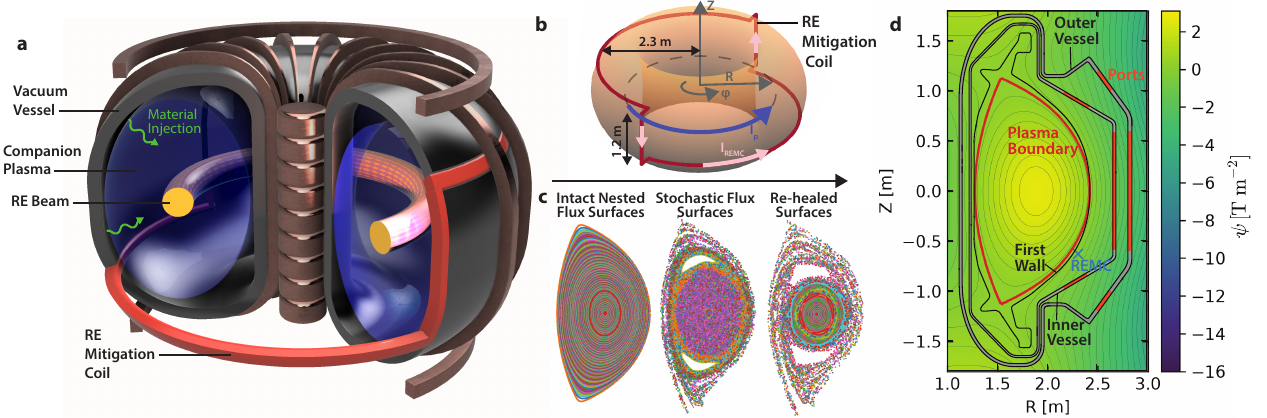}
\centering
\caption{
(a) A simplified diagram of a generic tokamak during a disruption, showing the RE beam (yellow), companion plasma (blue), the REMC (red), vacuum vessel (gray), and magnetic field coils (copper). The REMC is a relatively new, largely untested technology for RE mitigation on power-plant-scale tokamaks. (b) A simplified schematic of the SPARC `$n=1$' REMC. The REMC is positioned inside the vacuum vessel on the low-field side. The orange region illustrates the plasma volume. (c) Changes in the magnetic topology during a disruption, showing initially intact flux surfaces, flux surface stochasticization, and subsequent re-healing. (d) Magnetic flux distribution $\psi (\varphi = 0)$ at $t = \SI{0}{\milli\second}$ for the SPARC primary reference discharge, and the simulation geometry comprising the double-walled vacuum vessel (gray) with ports (orange). 
}
\label{fig:main}
\end{figure*}
 
 SPARC – a high-current, high-field tokamak designed to achieve a net fusion gain $Q >1$ \citep{creely2020overview} – will serve as a testbed for RE mitigation strategies on high-current machines \citep{sweeney2020mhd}. Baseline massive gas injection (MGI) on SPARC will use combined neon (Ne) and deuterium (D$_2$) injection to simultaneously mitigate thermal and RE loads \citep{kleiner2024extended,datta2026modeling,sweeney2026arc}. Simulations with the fluid-kinetic RE modeling framework DREAM \citep{hoppe2021dream}, however, indicate that relatively high deuterium densities ($n_D 
 {\gtrsim} 10^{22} \,\SI{}{\per\cubic\meter}$) are necessary to reduce the RE current below $\SI{1}{\mega\ampere}$ in SPARC disruptions \citep{ekmark2025runaway}. Moreover, a complete suppression of RE currents with material injection is generally considered difficult \citep{breizman2019physics}.

SPARC will therefore test a new technology – the RE Mitigation Coil (REMC) – which, as illustrated in \autoref{fig:main}\blue{(a-b)}, is a 3D conducting structure inside the vacuum vessel \citep{sweeney2020mhd,battey2023design} and represents a {\it passive} mechanism for RE suppression \citep{boozer2011two,tinguely2021modeling,izzo2022runaway,izzo2025disruption}. Currents induced in the REMC during the disruption will seed magnetic perturbations that generate MHD instabilities in the plasma volume \citep{boozer2011two,smith2013passive}. The nonlinear interaction of these REMC-driven MHD instabilities can break the initially intact magnetic flux surfaces (\autoref{fig:main}\blue{c}; left), producing stochastic magnetic topology (\autoref{fig:main}\blue{c}; center). Stochastic flux surfaces can suppress RE growth by enhancing the transport of REs out of the plasma volume \citep{rechester1978electron}, with successful mitigation achieved when the loss rate limits the RE generation rate \citep{smith2013passive,tinguely2021modeling,tinguely2023minimum}. This typically requires preventing intact flux surfaces from re-healing in the plasma (\autoref{fig:main}\blue{c}; right). 

While REMCs have been recently deployed on low-current tokamaks such as HBT-EP ($I_p {\leq} \SI{30}{\kilo\ampere}$) \citep{levesque2026first} and J-TEXT ($I_p {\leq} \SI{400}{\kilo\ampere}$) \citep{liu2025analysis}, and planned for medium-current machines such as TCV ($I_p {\leq} \SI{1.2}{\mega\ampere})$ \citep{Sheikh2025Runaway}, SPARC $(I_p = \SI{8.7}{\mega\ampere})$ will be the first to test the REMC in the high-current avalanche-dominated regime. Previous computational work that combines REMC-driven MHD activity (simulated using the MHD code NIMROD) \citep{izzo2022runaway} with RE transport (using the particle code ASCOT5 \citep{hirvijoki2014ascot}) and RE generation (using DREAM \citep{hoppe2021dream}) showed suppression of an otherwise unmitigated ${\sim}\SI{5}{\mega\ampere}$ RE beam predicted without the REMC \citep{tinguely2021modeling}. Recent NIMROD simulations, however, show significant re-healing of flux surfaces that can confine the RE population, raising concerns about the efficacy of the REMC \citep{izzo2022runaway,izzo2025disruption}. 

Here, we present 3D nonlinear extended MHD + RE fluid \citep{zhao2024simulation} simulations of RE mitigation using the REMC that complement, and improve upon, previous disruption modeling efforts for SPARC. We demonstrate that REMC-driven MHD activity significantly reduces the RE current from multi-MA levels to ${<}\SI{150}{\kilo\ampere}$ despite the re-healing of flux surfaces. The final termination of the RE beam is facilitated via scrape-off, which describes the opening-up of previously closed flux surfaces and the loss of plasma area once the plasma contacts the first wall due to vertical motion \citep{vallhagen2025reduced}. These simulations represent the first computational study that self-consistently combines 3D MHD activity, REs, and the REMC in full disruption simulations of a high-current tokamak. The simulations, performed with the extended MHD code {M3D-C1} \citep{jardin2012multiple}, offer several novel elements:
\begin{enumerate}
\item We self-consistently model RE generation and transport coupled with REMC-driven MHD activity, in contrast to previous efforts where different codes were used to separately model MHD activity, RE transport, and generation \citep{tinguely2021modeling,izzo2025disruption}, or where self-consistent MHD + RE fluid simulations of SPARC disruptions excluded the REMC \citep{datta2025coupled,datta2026modeling}.

\item We model the full mid-plane disruption in 3D, comprising the initial fast loss of stored thermal energy called the thermal quench (TQ), and the subsequent current decay stage called the current quench (CQ). Modeling the TQ in 3D enables us to incorporate changes in the plasma current (including an $I_p$-spike), safety factor, and RE population due to MHD activity, which is important for subsequent CQ evolution.

\item We include high-Z impurity and radiative corrections to RE avalanching, providing more realistic predictions of the dominant avalanching rate, which can be significantly higher for impurity-laden disruptions \citep{hesslow2019influence}. Furthermore, we include fusion-relevant primary sources such as tritium beta decay and Compton scattering \citep{datta2025coupled}, impurity-corrected Dreicer generation \citep{hesslow2019evaluation}, and prescribe a pessimistic, constant primary RE generation rate (${\sim}\SI{2} {\kilo\ampere}/\SI{}{\milli\second}$) during the TQ to mimic the hot-tail seed reported in previous work \citep{tinguely2021modeling,smith2008hot}.

\item We use a multi-region vacuum vessel geometry, comprising the first wall, inner and outer vessels with ports, and an outer vacuum region \citep{kleiner2024extended,datta2026modeling} (\autoref{fig:main}\blue{d}), as opposed to simplified vessel geometries used in previous modeling \citep{tinguely2021modeling,izzo2022runaway,datta2025coupled}. This multi-region geometry with realistic material-dependent resistivities allows us to self-consistently capture several important effects, such as resistive wall corrections to MHD instability growth rates and amplitudes, vertical motion, external magnetic flux diffusion, as well as currents in the vessel and the open-field-line `halo' region.

\item Lastly, we model the vertical displacement and final termination of the plasma and RE beam in 3D, as opposed to previous simulations, which either did not capture this important effect in elongated plasmas \citep{tinguely2021modeling,izzo2022runaway,izzo2025disruption,datta2025coupled} or modeled the vertical displacement event (VDE) in 2D \citep{datta2026modeling}.
\end{enumerate}

\begin{figure*}[t!]
\includegraphics[page=6,width=0.98\textwidth]{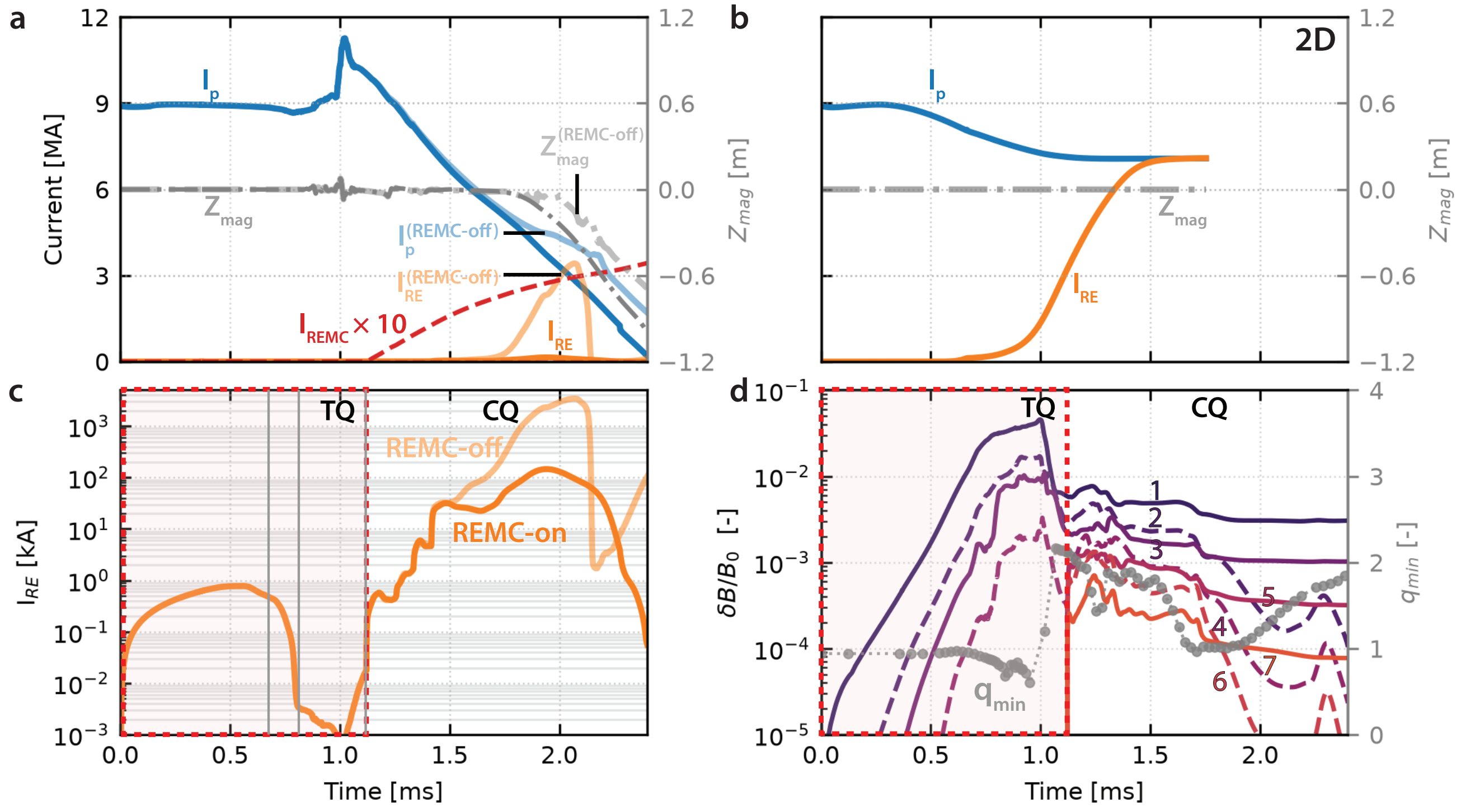}
\centering
\caption{
(a)~\textit{Left axis –} Temporal variation of the plasma current $I_p$ (blue), REMC current $I_{\text{REMC}}$ (red dashed) and the RE current $I_\text{RE}$ (orange) during the disruption. \textit{Right axis –} Vertical position of magnetic axis $Z_{mag}$ (gray) versus time. Semi-transparent lines show results from a simulation without the REMC. (b)~Plasma current (blue), RE current (orange) and magnetic axis vertical position $Z_{mag}$ (gray) in a 2D simulation without the REMC or 3D MHD instabilities. (c)~RE current $I_\text{RE}$ versus time on a logarithmic scale. (d)~\textit{Left axis –} Magnetic perturbation level $\delta B/B_0$ of different $n$ modes excited in the plasma in the REMC-on simulation. Here, $B_0\approx \SI{12.2}{\tesla}$ is the background magnetic field magnitude. Odd and even modes are represented by solid and dashed lines, respectively. $n < 4$ modes were resolved during the TQ, while $n < 8$ modes were resolved during the CQ. \textit{Right axis –} Variation of the minimum safety factor $q_{\text{min}}$.
}
\label{fig:current}
\end{figure*}

The initial equilibrium comprises the SPARC primary reference discharge, which describes a double-null diverted plasma with plasma current $I_p = \SI{8.7}{\mega\ampere}$ and fusion power $P_\text{fus} = \SI{140}{\mega\watt}$ \citep{rodriguez2020predictions}. \autoref{fig:main}\blue{d} shows the initial magnetic flux distribution for this equilibrium. We initialize the simulations with a pre-distributed population of roughly $\SI{3e21}{}$ Ne atoms, which radiatively cools the plasma during the TQ. Although baseline MGI on SPARC will comprise combined Ne + $D_2$ injection \citep{sweeney2020mhd}, the Ne-only case simulated here represents a {\it pathological} case where we expect fast current quench rates and large multi-MA RE plateaus, as seen in previous DREAM \citep{ekmark2025runaway}, DREAM + NIMROD \citep{tinguely2021modeling}, and M3D-C1 \citep{datta2026modeling} simulations.

\autoref{fig:main}\blue{b} shows the simplified geometry of the SPARC REMC, which comprises two toroidal sections (major radius $R_{coil} \approx \SI{2.3}{\meter}$, vertical position $Z_{coil} \approx \pm \SI{0.6}{\meter}$) positioned on the low-field (outboard) side of the vacuum vessel, connected via vertical legs. In this configuration, the REMC produces a dominant $n = 1$ toroidal harmonic, and other odd $n$ harmonics whose magnitudes fall with increasing $n$ \citep{tinguely2021modeling}. The 3D magnetic field for this REMC geometry is computed using a Biot-Savart model in {M3D-C1}, and the current response $I_{\text{REMC}}(t)$ is calculated from the solution of a lumped circuit model $L\text{d}_t I_{\text{REMC}} = -(M \text{d}_t I_\varphi + I_{\text{REMC}} \mathcal{R})$. Here, $L$ and $\mathcal{R}$ are the REMC's self-inductance and resistance respectively, and $M$ represents the mutual inductance between the REMC and total toroidal current $I_\varphi$ in the plasma and halo regions. The circuit model represents an improvement over previous simulations, which used a simple linear relationship between the plasma and REMC currents, and further assumed a maximum REMC current greater than the engineering limit (\SI{350}{\kilo\ampere}) for SPARC \citep{tinguely2021modeling,izzo2022runaway,izzo2025disruption}. 

\begin{figure*}[t!]
\centering
\includegraphics[page=2,width=1.0\textwidth]{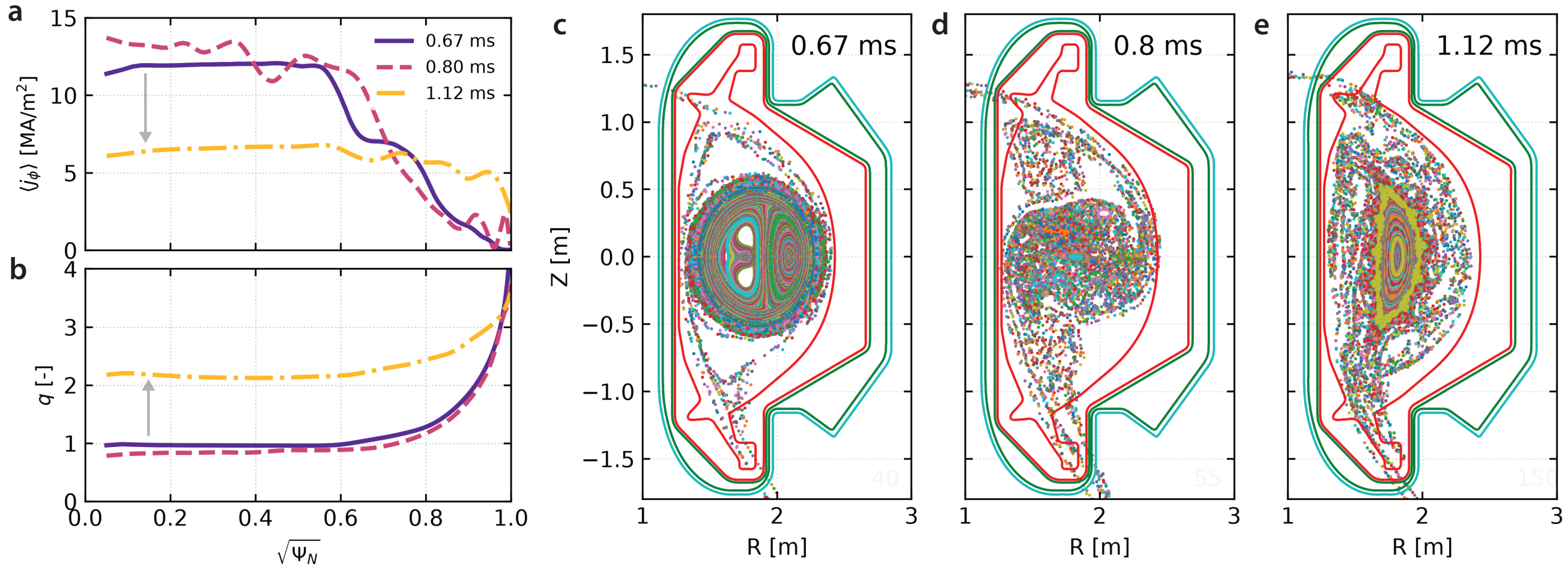}
 \caption{Evolution of (a) flux-averaged toroidal current density profile $j_\varphi(\sqrt{\psi_N})$ and (b) safety factor profile $q(\sqrt{\psi_N})$ during the TQ MHD event; and (c--e) Poincaré plots of the magnetic topology in the poloidal $R{\text{-}}Z$ plane during the TQ, corresponding to times indicated via solid gray vertical lines in \autoref{fig:current}\blue{c}.}
  \label{fig:DVV_TQ_MHD}
\end{figure*}

{\it Results –} \autoref{fig:current}\blue{a} shows the temporal evolution of the plasma current $I_p$ (blue), RE current $I_\text{RE}$ (orange), and REMC current $I_\text{REMC}$ (red dashed), with the RE current shown in more detail in \autoref{fig:current}\blue{c}. Here, $I_p$ includes both ohmic and RE currents inside the plasma boundary, but excludes any halo currents in the open-field-line region. A complete decay of the plasma current occurs over ${\sim}\SI{2.4}{\milli\second}$, and the RE current remains small, reaching a maximum value of about $\SI{146}{\kilo\ampere}$. The plasma current exhibits a large $I_p$-spike during the TQ, where $I_p$ rises to roughly $\SI{11}{\mega\ampere}$. As $I_p$ decays thereafter, the REMC current $I_{\text{REMC}}$ increases, reaching a maximum value of about $\SI{350}{\kilo\ampere}$. Instead of growing monotonically, the RE current $I_\text{RE}$ exhibits periods of decay or slowed growth, which, as shown later, coincide with periods of MHD instability growth. When $I_p$ falls below ${\sim}\SI{5}{\mega\ampere}$ later in the CQ, the plasma exhibits downward vertical motion. This is illustrated by the gray dashed line in \autoref{fig:current}\blue{a}, which shows the vertical position of the magnetic axis as a function of time. The final termination of the plasma and RE beam is facilitated via scrape-off, as the plasma contacts the first wall near the lower divertor.

In contrast, when we artificially suppress the REMC current during the disruption, a significantly larger RE current is obtained. These results are indicated via semi-transparent lines in \autoref{fig:current}\blue{(a,c)}. The RE current grows to roughly $\SI{3.4}{\mega\ampere}$, more than $20\times$ larger than that obtained with the REMC, before MHD activity during the VDE terminates the RE beam. These results, which are described in more detail in the following paragraphs, therefore indicate that the REMC plays a key role in limiting the RE current during the disruption.

\textit{Thermal Quench MHD and $I_p$-spike –} \autoref{fig:current}\blue{(d)} shows the magnetic perturbation level $\delta B / B_0$ of different toroidal $n$ harmonics excited in the plasma during the disruption. (Note that during the TQ and $I_p$-spike, no currents were allowed in the REMC, which assumes a REMC switch.) A broad spectrum of MHD modes with dominant $n = 1$ activity is excited during the TQ, with peaks in the mode perturbation level between $t \approx 0.8 \text
{ - } \SI{1.1}{\milli\second}$. This coincides with the $I_p$-spike seen in \autoref{fig:current}\blue{a}. \autoref{fig:current}\blue{d} shows the evolution of the minimum safety factor $q_{\text{min}}$ (gray markers) overlaid on the $\delta B/B_0$ evolution. In a tokamak, the safety factor $q(r) \equiv B_\varphi r/B_p R$, which is a function of the toroidal and poloidal magnetic fields ($B_\varphi$ and $B_p$ respectively) and the major and minor radii ($R$ and $r$ respectively), represents the stability of the plasma to MHD instabilities.  As observed in \autoref{fig:current}\blue{d}, $q_{\text{min}}$ drops from roughly $0.93$ to about $0.75$ initially, but jumps to about 2 by  $t \approx \SI{1.1}{\milli\second}$, which marks the end of the TQ MHD event.

\begin{figure*}[t!]
\includegraphics[page=3,width=1.0\textwidth]{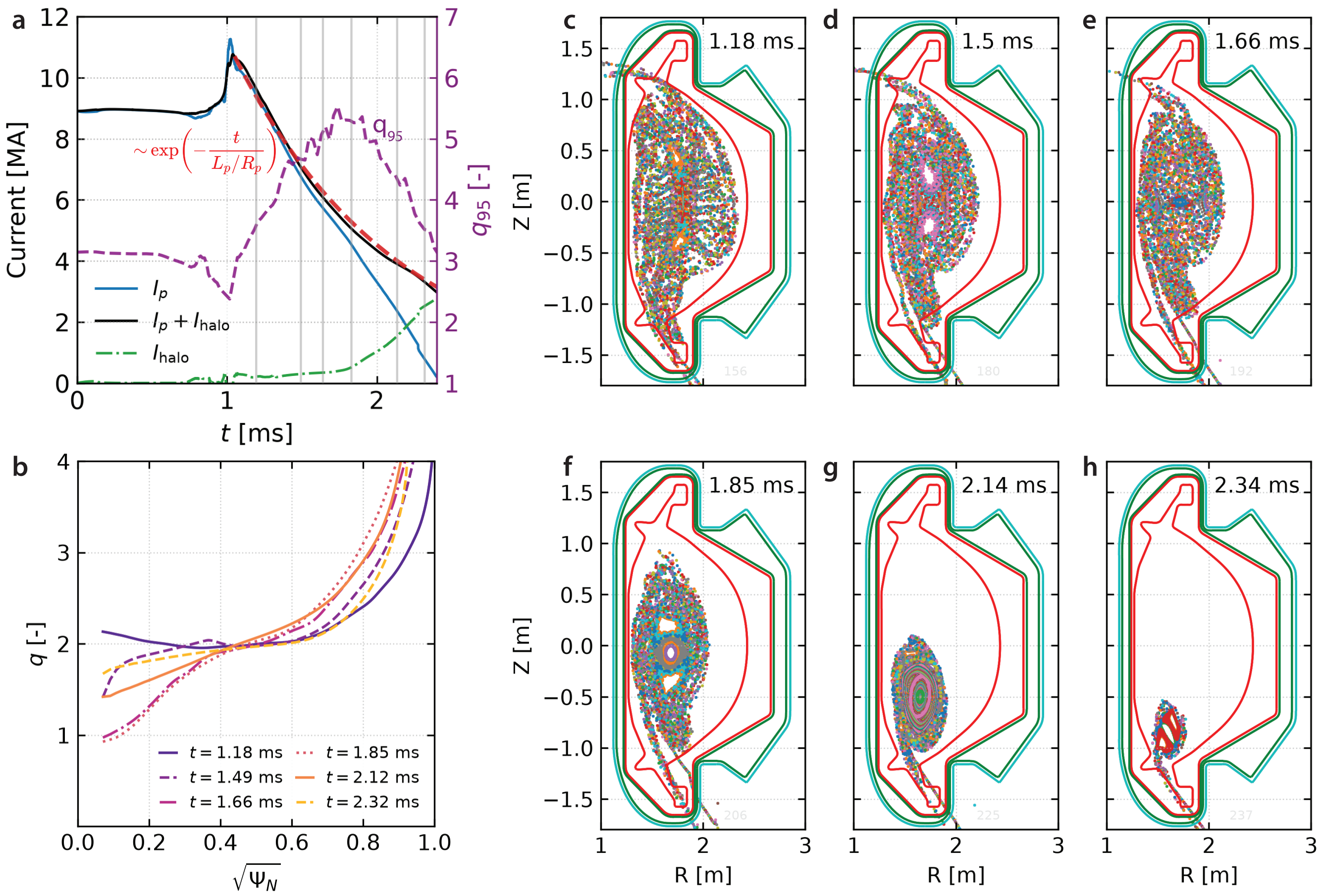}
    \caption{(a) Plasma current $I_p$ (blue) and (toroidal) halo current $I_{\text{halo}}$ (green dashed) in the REMC-on simulation. The red dashed line represents exponential current decay with the characteristic time $\tau_\text{CQ} \approx L_p/R_p$ in the post-TQ plasma. [right axis] Evolution of the edge safety factor $q_{95}$ in the REMC-on simulation (purple).
    (b) Safety factor profile $q(\sqrt{\psi_N})$ at various times during the CQ; and 
    (c--h) Poincaré plots of the magnetic topology during the CQ and VDE in the REMC-on simulation, corresponding to times indicated via gray solid vertical lines in (a).}
  \label{fig:DVV_CQ_MHD}
\end{figure*}

MHD activity causes significant flattening of the current density profile $j_\varphi(\sqrt{\psi_N})$. As shown in \autoref{fig:DVV_TQ_MHD}\blue{a},  the on-axis current density falls from about $j_\varphi \approx \SI{12}{\mega\ampere\per\meter\squared}$ to roughly $\SI{6}{\mega\ampere\per\meter\squared}$ at $t \approx \SI{1.12}{\milli\second}$, while the safety factor profile $q(\sqrt{\psi_N})$ rises, as shown in \autoref{fig:DVV_TQ_MHD}\blue{b}, consistent with the observed flattening of current density. The  internal inductance $L_i$ of the plasma decreases from about $\SI{1.24}{\micro\henry}$ initially to $\SI{0.62}{\micro\henry}$ at $t \approx \SI{1.1}{\milli\second}$, acting as a driver for the $I_p$-spike. \autoref{fig:DVV_TQ_MHD}\blue{(c-e)} shows Poincaré plots of the magnetic topology during the TQ. Tearing activity produces $(m,n) = (3,2)$, (4,3), (2,1) and (3,1) magnetic islands in the outer $q > 1$ region, breaking up flux surfaces via island overlap, while a (1,1) internal kink appears in the inner $q < 1$ region (\autoref{fig:DVV_TQ_MHD}\blue{c}). As seen in \autoref{fig:DVV_TQ_MHD}\blue{d}, the magnetic topology becomes almost fully stochastic by $t \approx \SI{0.8}{\milli\second}$. By $t \approx \SI{1.12}{\milli\second}$, however, intact flux surfaces reform around the magnetic axis (\autoref{fig:DVV_TQ_MHD}\blue{e}), consistent with a decay in the magnetic perturbation level and a flattening of the safety factor profile after the MHD event. While the RE current grows linearly during the initial part of the TQ, as shown in \autoref{fig:current}\blue{c}, $I_\text{RE}$ decreases rapidly from ${\sim}\SI{1}{\kilo\ampere}$ to ${\sim}\SI{1}{\ampere}$ between $t \approx 0.6 \text{ - } \SI{1.1}{\milli\second}$, consistent with the onset of magnetic stochasticity due to TQ MHD activity. These results are consistent with previous NIMROD simulations, which also showed a rise in $q_{\text{min}} > 2$ accompanying the $I_p$-spike, together with strong losses of RE test particles during the TQ \citep{izzo2022runaway,izzo2025disruption}.


\textit{Current Quench –} After the TQ, the maximum electron temperature $T_\text{{e,max}}$ falls to ${\sim}\SI{10}{\electronvolt}$, and remains at this low value for the rest of the simulation, providing an inductive $L_p/R_p$ time of about $\tau_\text{CQ} \approx \SI{1}{\milli\second}$. \autoref{fig:DVV_CQ_MHD}\blue{a} shows the plasma and halo current evolution in the REMC-on simulation. During the initial part of the CQ, $I_p$ decays at a rate of roughly $\text{d}_t{I}_p \approx \SI{-6}{\mega\ampere}/\SI{}{\milli\second}$. As shown in \autoref{fig:DVV_CQ_MHD}\blue{a}, the decay rate of the total toroidal current in the plasma-halo region is consistent with the expected exponential decay rate with characteristic time $\tau_\text{CQ}$ (red dashed line).

A broad spectrum of MHD modes is obtained during the CQ in the REMC-on simulation, as shown in \autoref{fig:current}\blue{d}, but $n = 1$ modes remain the dominant harmonic. $\delta B/B_0$ exhibits several peaks during the CQ, associated with the periodic growth and decay of MHD instabilities (\autoref{fig:current}\blue{d}). During the early CQ, MHD activity is accompanied by oscillations of $q_\text{min}$ about 2. \autoref{fig:DVV_CQ_MHD}\blue{(c-h)} shows the magnetic topology at various times during the CQ. Re-healed flux surfaces originally observed around the magnetic axis after TQ MHD activity (\autoref{fig:DVV_TQ_MHD}\blue{e}) are re-stochasticized by successive MHD activity, and we recover fully stochastic surfaces again by $t \approx  \SI{1.18}{\milli\second}$ (\autoref{fig:DVV_CQ_MHD}\blue{c}). Periodic formation and re-stochasticization of $(2,1)$ magnetic islands is also observed during this time (\autoref{fig:DVV_CQ_MHD}\blue{d}). Intact surfaces initially present around the magnetic axis early in the CQ (\autoref{fig:DVV_TQ_MHD}\blue{e}) enable a rapid increase in the RE current, as shown in \autoref{fig:current}\blue{c}. By $t \approx \SI{1.3}{\milli\second}$, $I_p$ has fallen to its pre-disruption value, and the RE current has increased to $I_\text{RE} \approx \SI{3}{\kilo\ampere}$. The re-stochasticization of flux surfaces reduces the net RE growth rate, and also produces periods of intermittent RE losses, as seen in \autoref{fig:current}\blue{c}.

 \begin{figure*}[t!]
\includegraphics[page=4,width=0.98\textwidth]{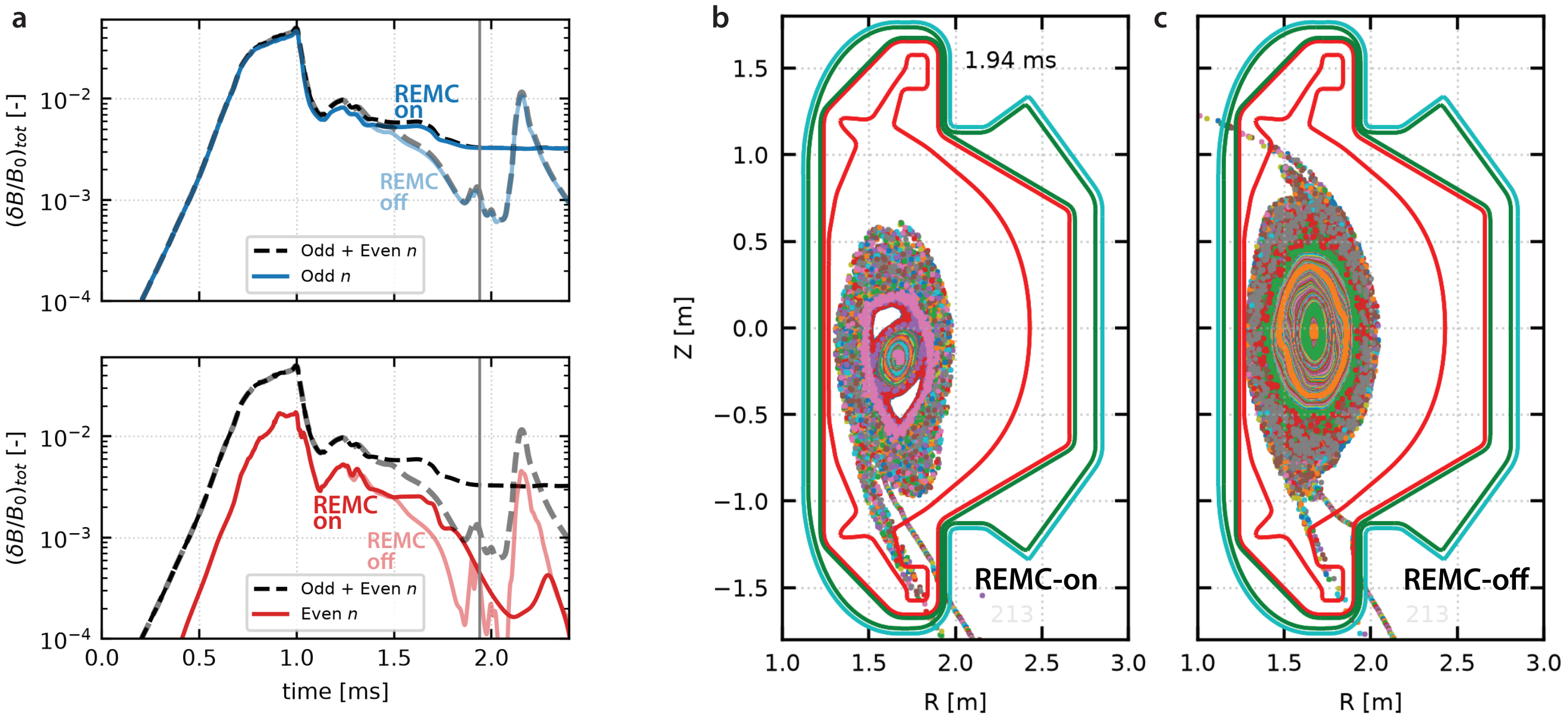}
\caption{
(a) Comparison of total magnetic perturbation level of odd $n$ (blue, top) and even $n$ (red, bottom) toroidal modes in simulations with (opaque) and without (semi-transparent) the REMC. The total perturbation level of odd and even $n$ modes is indicated by black dashed lines in both panels. (b-c) Comparison of the magnetic topology at $t \approx \SI{1.94}{\milli\second}$ (gray vertical line in a) in the REMC-on and REMC-off simulations. 
}
\label{fig:comparison}
\end{figure*}

Between $t \approx 1.5\text{ - }\SI{1.8}{\milli\second}$, $q_\text{min}$ falls rapidly, dropping slightly below unity. This triggers further MHD activity, producing peaks in $\delta B/B_0$ (\autoref{fig:current}\blue{d}), increased magnetic stochasticity (\autoref{fig:DVV_CQ_MHD}\blue{e}), and RE losses (\autoref{fig:current}\blue{c}) during this period. The evolution of the q-profile at different times during the CQ is shown in \autoref{fig:DVV_CQ_MHD}\blue{b}. The initially flat q-profile becomes increasingly steeper, bringing flux surfaces in the inner region into resonance with the dominant $n=1$ perturbations driven by the REMC. The change in the q-profile is driven primarily by a peaking of the current density caused by a lower resistivity in the hotter, inner plasma region. Current re-distribution by MHD activity causes a subsequent rise in $q_{\text{min}}$ after $t \approx \SI{1.8}{\milli\second}$, and the q-profile becomes increasingly flatter (\autoref{fig:DVV_CQ_MHD}\blue{b}). This flattening produces a region of sparse resonant flux surfaces, limiting overlap between modes on neighboring flux surfaces, and enabling re-healing around the magnetic axis, as shown in \autoref{fig:DVV_CQ_MHD}\blue{(f-g)}. Similar re-healing around the magnetic axis has been reported previously in REMC simulations \citep{smith2013passive,izzo2022runaway,izzo2025disruption}. Re-healing is accompanied by a strong increase in the RE current to roughly $I_\text{RE} \approx \SI{146}{\kilo\ampere}$ between $t \approx \SI{1.7}{\milli\second} \text{ - } \SI{1.95}{\milli\second}$.


\textit{Vertical motion \& termination –} The magnetic topology during vertical motion of the plasma in the REMC-on simulation is visualized in \autoref{fig:DVV_CQ_MHD}\blue{(f-h)}. The plasma begins to move vertically downwards around $t \approx \SI{1.8}{\milli\second}$ and contacts the first wall near the lower divertor. This post-TQ plasma motion constitutes a ``cold'' VDE \citep{boozer2019halo}, triggered by the plasma current falling below a critical value, which for SPARC is roughly $\SI{5}{\mega\ampere}$ \citep{clauser2024modeling,datta2026modeling}.  The plasma cross-sectional area shrinks during this time, at a rate of roughly $\text{d}_t{A}_p\approx\SI{-2}{\meter \squared}/\SI{}{\milli\second}$, accompanied by an increase in the halo current, as current is lost from within the plasma boundary to the open-field-line region (\autoref{fig:DVV_CQ_MHD}\blue{a}). 

During vertical motion, re-healed flux surfaces exist around the magnetic axis, while outer flux surfaces remain stochastic (\autoref{fig:DVV_CQ_MHD}\blue{g}). The region of intact surfaces, however, shrinks with time. As seen in \autoref{fig:current}\blue{c}, the RE current increases during the initial part of the VDE ($t \approx \SI{1.8}{} \text{ - } \SI{1.95}{\milli\second}$), before decreasing afterwards. The minimum  safety factor $q_\text{min}$ continues to rise during this time, and the q-profile, as noted earlier, becomes increasingly flat near the magnetic axis (\autoref{fig:DVV_CQ_MHD}\blue{b}). In contrast, the edge safety factor $q_{95}$, which increases prior to vertical displacement, starts to decrease, as shown in \autoref{fig:DVV_CQ_MHD}\blue{a}. The change in $q_{95}$ is consistent with the expected scaling $q_{95}\propto a^2/I_p$, showing a decrease when the cross-sectional area begins to fall faster than the plasma current (that is, $\dot{q}_{95}/q_{95} {\sim}[2\dot{a}/a-\dot{I}_p/I_p]$ becomes negative). 3D MHD instabilities grow, exhibiting peaks in $\delta B/B_0$ (primarily even $n$ modes) around $t \approx \SI{2.3}{\milli\second}$ (\autoref{fig:current}\blue{d}). MHD activity also produces an acceleration of the RE loss rate, as seen in \autoref{fig:current}\blue{c}.

\textit{Comparison with the REMC-off case –} A comparison of the results obtained with and without the REMC demonstrates important coupled dynamics of flux surface re-healing, vertical displacement, and RE generation. \autoref{fig:comparison}\blue{a} compares the total magnetic perturbation $(\delta B/B_0)_{tot}$ of odd $n$ (blue) and even $n$ (red) modes obtained in simulations with and without the REMC. In the early CQ ($t \lesssim \SI{1.4}{\milli\second}$), the magnetic perturbation level and RE current evolution (\autoref{fig:current}\blue{c}) are identical in both cases. This indicates that during this period, when the REMC current is small ($I_\text{REMC} \lesssim \SI{100}{\kilo\ampere}$), MHD activity grows primarily from perturbations pre-existing in the plasma from the TQ, instead of being actively seeded by the REMC. Thereafter, the magnetic perturbation level becomes higher with the REMC than in the REMC-off case. The total magnetic perturbation remains high with the REMC primarily because of odd $n$ modes actively driven by the coil. This is observed in \autoref{fig:comparison}\blue{a}, which shows that the $\delta B/B_0$ of even $n$ modes decays significantly later in the CQ ($t \gtrsim \SI{1.5}{\milli\second}$), while that of odd $n$ modes remains high. In contrast, the REMC-off case demonstrates a strong decay of both odd and even $n$ modes, consistent with the absence of any actively driven perturbations in this simulation. The lower perturbation level obtained without the REMC results in earlier and more significant re-healing of flux surfaces, as shown in \autoref{fig:comparison}\blue{(b-c)}, which compares the magnetic topology between the two cases at $t \approx \SI{1.94}{\milli\second}$. This is consistent with the earlier and significantly faster RE current growth between $t \approx \SI{1.5}{} \text{ - }\SI{2.1}{\milli\second}$ observed without the REMC in \autoref{fig:current}\blue{c}.

The substantial re-healing of flux surfaces without the REMC enables the RE population to grow and become the primary current carrier by $t \approx \SI{2}{\milli\second}$. The higher $I_\text{RE}$ slows down the plasma current decay rate, in turn, slowing down the vertical motion \citep{boozer2019halo,clauser2021iter}. This is illustrated in \autoref{fig:current}\blue{a}, which shows that the plasma current $I_p$ becomes higher and the vertical displacement is delayed in the simulation without the REMC, providing a feedback mechanism that enables the RE current to grow to a larger value without significant scrape-off losses. A similar modification of the VDE by RE currents was reported previously in 2D MHD simulations \citep{datta2026modeling}.

Finally, the RE current exhibits a rapid loss event (\autoref{fig:current}\blue{c}), associated with a large peak in the $\delta B/B_0$ (\autoref{fig:comparison}\blue{a}), around $t \approx \SI{2.15}{\milli\second}$ in the simulation without the REMC, followed by eventual re-avalanching of the RE current. An inspection of the safety factor shows that the q-profile becomes non-monotonic due to off-axis peaking of the current profile \citep{datta2025coupled}. This, in turn, triggers double tearing modes, primarily at the $2/1$ and $3/1$ resonant surfaces, causing strong MHD activity and rapid stochasticization of flux surfaces. Consequently, a RE plateau (i.e. a freezing of the $I_p$ decay due to the RE current) is not obtained; instead, a brief "shoulder" in the plasma current evolution is produced (\autoref{fig:current}\blue{a}). Similar qualitative behavior was also observed in early disruption experiments of diverted plasmas on the JET tokamak, and postulated to be a consequence of reduced RE confinement in diverted plasmas  \citep{harris1990comparisons}.


{\it Discussion –} Our results show that while a complete suppression of RE currents may be difficult to achieve on high-current tokamaks, the REMC may still provide a viable pathway for RE mitigation, even in {\it pathological} situations, as seen in these simulations. The re-healing of flux surfaces due to decaying magnetic perturbation magnitudes and a flattening of the q-profile following MHD activity enables a rapid rise of the RE current in the avalanche-dominated regime. However, by delaying the onset of re-healing and limiting the extent of intact flux surfaces, REMC-driven MHD activity is likely to limit the RE current to a level where vertical displacement of the plasma also contributes to RE losses during the disruption.

Recent work has explored the role of vertical displacement in limiting the avalanche gain in high-current tokamaks through reduced-order modeling of  VDE dynamics \citep{vallhagen2025reduced,martin2026simplified}. For vertical displacement to contribute to limiting the RE current, it is necessary that the RE current remains sufficiently low prior to vertical motion. In cases where an appreciable RE current develops, vertical motion can be slowed down (as shown in \autoref{fig:current}\blue{a}), and even completely avoided if a RE plateau maintains the plasma current above the critical value. This was observed when the simulations were repeated in 2D (without the REMC or 3D MHD instabilities); here, a large RE plateau of $\SI{7}{\mega\ampere}$, above the critical value, is obtained, and vertical motion is suppressed (see \autoref{fig:current}\blue{b}). A similar suppression of the "cold" VDE due to large RE plateaus was also observed in some 3D SPARC disruption simulations performed previously \citep{datta2026modeling}.

These simulations represent the first high-fidelity study of disruptions on a high-current tokamak that incorporates coupled feedback between RE physics, 3D MHD instabilities, and the REMC, together with various important disruption-relevant phenomena, such as TQ MHD activity, the I$_p$-spike, impurity effects, vertical motion, halo currents, and induced vessel currents. As noted earlier, while material injection – a RE mitigation strategy widely tested on present-day devices – can reduce RE generation rates, high injection densities and assimilation rates, which may be challenging to achieve, are necessary \citep{sweeney2026arc}, making the REMC a valuable mechanism for mitigating REs in avalanche-dominated high-current machines. In addition to providing predictive capabilities for RE mitigation with the REMC, this work also provides valuable data for making predictions of energy deposition on PFCs, heat loads, and electromagnetic forces that inform engineering requirements for disruption-robust, power-plant-scale tokamaks.

\setcitestyle{square}
\bibliography{main}
\setcitestyle{square}

\pagebreak

\section{Methods}

{\bf Simulation Setup.} The initial equilibrium comprises the primary reference discharge (PRD) in SPARC, which describes a double-null diverted plasma with plasma current $I_p = \SI{8.7}{\mega\ampere}$, toroidal magnetic field $B_0 = \SI{12.2}{\tesla}$, and volume-averaged ion density and electron temperature of $\langle n_i \rangle \approx 3\times 10^{20} \SI{}{\per \meter \cubed}$ and $\langle T_e \rangle \approx \SI{7}{\kilo \electronvolt}$ respectively \citep{rodriguez2020predictions}. We note that the SPARC PRD equilibrium exhibits a $q =1$ flux surface at roughly half the minor radius. While the SPARC PRD comprises a deuterium-tritium plasma, we model the main ion species as purely deuterium in our simulations for simplicity.

The simulation geometry comprises the first wall, inner vessel, outer vessel, and an outer vacuum region, with realistic resistivity values specified in each region ($\eta \approx \SI{8e-7}{\ohm\meter}$ in the inner/outer vessel, non-uniform parallel Spitzer resistivity in the plasma). The boundary of  the outer vacuum region marks the boundary of the simulation domain and is treated as a perfect conductor. This conducting boundary is placed far away from the vessel boundaries, so as to not significantly affect the dynamics. Ports in the inner and outer vessel walls, which provide diagnostic access, are also included, and are modeled as regions of high resistivity in the toroidal direction ($\eta_\varphi \approx 10^{-3}\SI{}{\ohm\meter}$), and low resistivity in the poloidal direction ($\eta_{RZ} \approx \SI{8e-7}{\ohm\meter}$). The domain is discretized with an unstructured mesh comprising triangular finite elements in the (R, Z) plane and prism-shaped elements in the toroidal direction. {M3D-C1} solves the MHD equations in the region within the first wall using reduced quintic $C^1$ shape functions in the (R,Z) plane and cubic polynomial shape functions in the toroidal direction \citep{jardin2012multiple}. Details about the finite element treatment and numerical schemes used in M3D-C1 are provided in \blue{Ref.}~\citenum{jardin2012multiple}  and \blue{Ref.}~\citenum{ferraro2009calculations}.

The Ne ionization level and emitted radiation (which comprises line, bremsstrahlung, and recombination radiation) are calculated using the KPRAD module in {M3D-C1} \citep{ferraro20183d}. Once ionized, Ne is assumed to be at the same temperature $T_i$ as the primary ion species. Energy is removed from the electron temperature to account for the energy required to bring the ionized Ne up to the ion temperature. Here, the injected Ne is pre-distributed in the plasma; preliminary simulations with Ne injected at MGI valve locations performed with a simplified geometry showed that the injected Ne becomes well-distributed in the plasma during the TQ. Finally, we note that baseline MGI on SPARC will comprise combined Ne + D$_2$ injection; this is expected to slow down the CQ time (thus, slowing down the REMC ramp-up rate) as well as the RE generation rate, as seen in previous 2D simulations which showed a maximum RE current of about $\SI{2}{\mega\ampere}$ for the SPARC baseline MGI scenario  \citep{datta2026modeling}.

Both the TQ and CQ are modeled in 3D. The TQ is modeled with 4 toroidal planes (resolving $n < 4$ toroidal modes), while the CQ is modeled with 8 toroidal planes (resolving $n < 8$ modes). The chosen toroidal resolution is sufficient to resolve the dominant $n$ harmonics excited in the system. Preliminary simulations performed with higher toroidal resolution (resolving $n < 12$) showed similar results. The primary goal of modeling the TQ in 3D is to enable MHD instabilities to redistribute the thermal pressure, current density, and impurity density, providing more practical profiles at the start of the CQ. This represents an improvement over previous modeling efforts \citep{tinguely2021modeling,datta2026modeling}, where the TQ was modeled in 2D, and changes to the current profile and safety factor due to 3D MHD instabilities were therefore not included. To avoid modeling a long pre-TQ phase (which adds computational cost), the isotropic thermal diffusivity $\chi$ was initially increased to about ${\sim}10^3\SI{}{\meter \squared \per \second}$ during the TQ phase of the disruption to bring the temperature down to several-hundred $\SI{}{eV}$. $\chi$ was then reduced to $\chi {\sim}\SI{10}{\meter \squared \per \second}$ by the end of the TQ. 

{\bf RE Fluid Model.} The RE fluid model \citep{zhao2024simulation,datta2025coupled} in M3D-C1 is used to self-consistently calculate RE generation and transport during the disruption. Details of the RE fluid model can be found in \blue{Ref.} \citenum{zhao2024simulation}. Secondary avalanche generation is calculated using the modified Rosenbluth-Putvinski formula proposed by \citeauthor{hesslow2019influence}, which includes the effect of high-Z impurities as well as synchrotron and bremsstrahlung radiation on the effective critical electric field and the avalanche rate, as obtained in kinetic simulations \citep{hesslow2019influence}. Primary RE generation via the hot-tail mechanism is expected to be significant in next-generation tokamaks \citep{smith2008hot}. The hot-tail seed is very sensitive to the TQ physics, and a direct implementation of hot-tail generation, which often requires a kinetic treatment, is not presently available in M3D-C1. Instead, we prescribe a constant primary generation rate during the TQ, providing a seed of roughly $I_{RE} = \SI{2}{\kilo\ampere}$, to mimic the hot-tail seed reported by \blue{Ref.}~\citenum{tinguely2021modeling} in previous simulations. We note that this value likely represents an upper bound on the hot-tail seed, consistent with predictions made using analytical models assuming fast TQ times ${\sim}\SI{0.1}{\milli\second}$ \citep{smith2008hot}. Activated sources – Compton scattering (computed using OpenMC simulations of the photon flux inside the SPARC vacuum vessel from activated walls) and tritium beta decay (assuming half the primary ion density is composed of tritium) – are also included in the present simulations \citep{datta2025coupled}. Dreicer generation (both classical \citep{dreicer1959electron} and impurity-corrected \citep{hesslow2019evaluation}) is also available in M3D-C1 \citep{datta2025coupled,datta2026modeling}; here we use the impurity-corrected Dreicer model implemented as a neural network to calculate the Dreicer generation rate.

The RE fluid model used in our simulations provides the capability to tractably incorporate RE generation and transport in MHD simulations; however, a fluid treatment excludes information about electron energy, pitch angle with respect to the local magnetic field vector, and particle gyro-motion. RE losses can be sensitive to these quantities; for instance, high-energy electrons may be preferentially lost to the wall due to large drifts and gyro-orbits. MHD simulations have explored kinetic coupling \citep{puel20263d,liu2022thermal}; however, a direct coupling of kinetic REs with extended MHD codes for full disruption simulations remains challenging. Differences in RE transport and losses between the fluid model and those calculated by particle codes initialized with magnetic fields exported from M3D-C1 will be the subject of a future investigation.

{\bf REMC Model.} The time-resolved REMC current is calculated by solving the lumped circuit equation with self-inductance $L = \SI{19}{\micro\henry}$, resistance $R =  \SI{1.29}{\milli\ohm}$ and effective mutual inductance $M = \SI{0.94}{\micro\henry}$. These values are obtained from 3D electromagnetic modeling of the REMC using the code ThinCurr  \citep{battey2023design}. The self-inductance and resistance of the coil are geometry- and material-dependent constants; the mutual inductance can change during the disruption, especially during vertical motion when the plasma magnetic axis and current centroid are displaced. A simple analytical estimate of the change in the mutual inductance $M$ between the REMC and the plasma due to changes in the current centroid $Z_c$ can be obtained by using thin coil approximations for the REMC and plasma components, and considering the fluxes through the upper and lower regions enclosed by the REMC contour (see \autoref{fig:main}\blue{b}) \citep{rosa_mutual_1906}. For $|Z_c| \leq \SI{1}{\meter}$, the mutual inductance is expected to vary by $< 10\%$, representing a small change in the inductive coupling. We further note that while the plasma magnetic axis is displaced by $Z_c \approx \SI{1}{\meter}$ during the VDE in our simulation, the change in the current centroid is expected to be lower due to halo current contributions.

\begin{figure*}[b!]
\includegraphics[page=5,width=1.0\textwidth]{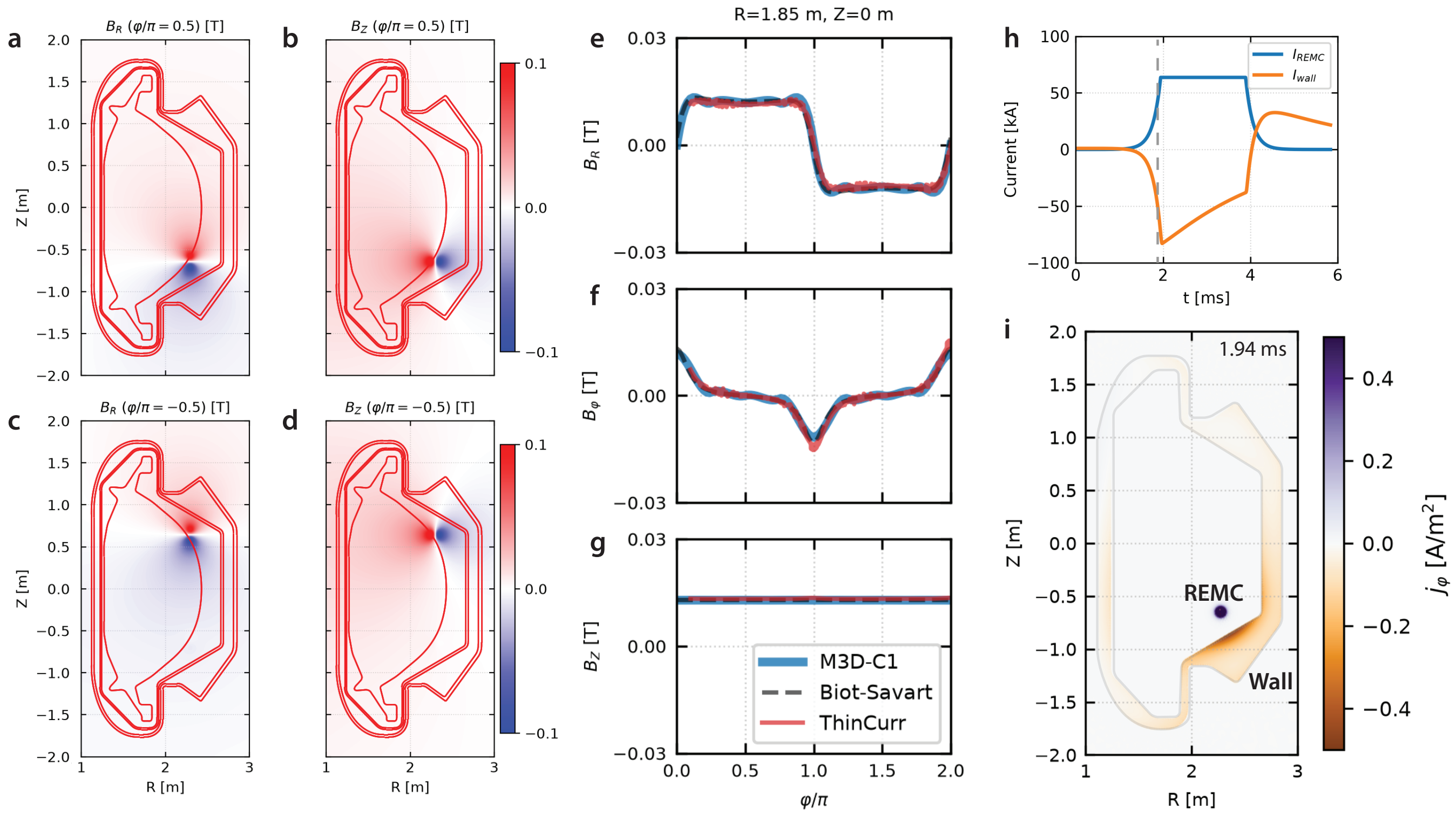}
\caption{
(a-b) The radial $B_R$ and vertical $B_Z$ components of the REMC magnetic field at toroidal position $\varphi = \pi/2$ for a REMC current of \SI{50}{\kilo\ampere}. (c-d) $B_R$ and $B_Z$ at $\varphi = -\pi/2$. (e-g) Variation of the REMC magnetic field components $B_R,\, B_\varphi$ and $B_Z$  with toroidal angle $\varphi$ at $(R,Z) = (\SI{1.85}{\meter}, \, \SI{0}{\meter})$. Blue lines are the M3D-C1 calculation, dashed black lines are an independent 3D Biot-Savart solution, and the red lines are ThinCurr results. (h) Induced currents in the wall (orange) due to an artificially prescribed REMC current (blue) comprising ramp-up, flat, and ramp-down phases. A high resistivity was prescribed in the plasma region to suppress currents in this region for this example. Note that this example used a single contiguous wall for simplicity, unlike the double-walled vessel used in the simulations described in this manuscript. (i) Toroidal current density showing wall currents during the REMC ramp-up phase [vertical dashed line in (h)]. Negative currents are induced in the wall during this phase. 
}
\label{fig:extended_data}
\end{figure*}

A Biot-Savart model calculates the magnetic field produced by the REMC in M3D-C1, and includes contributions from both the toroidal sections and the legs of the coil. \autoref{fig:extended_data}\blue{(a-d)} show the vertical and radial components of the REMC magnetic field at $\varphi=\pm \pi/2$ for $I_{\text{REMC}} = \SI{50}{\kilo\ampere}$, which are the dominant contributions produced by the toroidal sections of the coil. The field computed by M3D-C1 is in good agreement with that calculated by an independent 3D Biot-Savart solver (within $5\%$) and by ThinCurr (within $7\%$). This is shown in \autoref{fig:extended_data}\blue{(e-g)}, which shows the toroidal variation of $B_R$, $B_\varphi$, and $B_Z$ at the original magnetic axis position.

The REMC's 3D non-axisymmetric magnetic field ${\bf B}$ couples with the MHD equations via the scalar variables $\psi$ [where $(B_R,\,0,\, B_Z) = \nabla \psi \times \nabla \varphi$] and $\mathcal{I}\equiv RB_\varphi$ \citep{ferraro2009calculations}, and via the time-varying terms $\text{d}_t{\psi}_{\text{REMC}}, \text{d}_t{\mathcal{I}}_{\text{REMC}} \propto \text{d}_t{I}_{\text{REMC}}$, which drive induced currents in the plasma and surrounding conducting structures. The Biot-Savart magnetic field calculation is used to construct the flux $\psi_\text{{REMC}}$ and axial field terms $\mathcal{I}_\text{{REMC}}$, which are evolved in M3D-C1 instead of the magnetic field components and preserve the divergence-free condition $\nabla \cdot {\bf B} = 0$ \citep{ferraro2009calculations}. The scalar quantities couple to the MHD equations, as described in \blue{Ref.}~\citenum{ferraro2009calculations}. In addition, the time-derivatives $\text{d}_t{\psi}_\text{{REMC}}$ and $\text{d}_t{\mathcal{I}}_\text{{REMC}}$ are added to the flux and axial field evolution equations, and induce an electromotive force in the plasma and surrounding conducting structures that screen the REMC's time-varying field. As a simplified example demonstrating this effect, we show a simulation in \autoref{fig:extended_data}\blue{(h-i)} of a REMC-wall system, where the REMC current is first artificially ramped up to $I_\text{{REMC}} \approx \SI{60}{\kilo\ampere}$, then held constant, and finally ramped down (shown by the blue curve in \autoref{fig:extended_data}\blue{h}). Note that this example used a single contiguous wall for simplicity, unlike the double-walled vessel used in the simulations described in this manuscript. During the ramp-up phase, negative currents are induced in the wall, as expected from Lenz's law. \autoref{fig:extended_data}\blue{i} shows the current density in the wall during the ramp-up phase; strong negative currents are induced in the wall close to the REMC position. When the REMC current is held constant, the wall current decays resistively, and finally, when the REMC current is ramped down, positive currents are induced in the wall (see \autoref{fig:extended_data}\blue{h}). A similar response is produced in the plasma when induced currents are allowed in the plasma region (not shown here).

{\bf Computational Cost.} The simulations presented in this work were performed on the NERSC Perlmutter CPU partition, and took roughly 10,000 node hours (128 cpus per node).

\section{Data Availability}

The data that support the findings of this study are available upon reasonable request from the authors.

\section{Code Availability}

M3D-C1 is open-source software, publicly available at \blue{https://github.com/PrincetonUniversity/M3DC1.git}.

\section{Competing Interests}

The authors declare no competing interests.

\section{Acknowledgments}

This work is supported by Commonwealth Fusion Systems (RPP021: Disruptions). The simulations presented in this paper  used resources of the National Energy Research Scientific Computing Center (NERSC) [10.13039/100017223], a Department of Energy Office of Science User Facility (m3195 and m4829 for year 2025-26).  This work is also supported by the U.S. Department of Energy under contract numbers DE-AC02-09CH11466 (MiRACL) and DE-SC0026438 (CTC). The authors acknowledge J. Chen (PPPL), C. Liu (Peking University), and E. S. Seol (Rensselaer Polytechnic Institute) for their continued support. The authors also thank J. Boguski (CFS) for help with the REMC lumped circuit model.

\end{document}